\documentclass[
superscriptaddress,
reprint,
 amsmath,amssymb,
 aps,pra,
floatfix,
]{revtex4-2}
\usepackage{graphicx}
\usepackage{dcolumn}
\usepackage{bm}
\usepackage{braket}
\usepackage{float}
\usepackage[breaklinks=true,colorlinks,citecolor=blue,linkcolor=blue,urlcolor=blue]{hyperref}

\begin{document}
\preprint{APS/123-QED}
\title{Study on quantum thermalization from thermal initial states in a superconducting quantum computer}
\author{Marc Espinosa Edo}
\email{marc.espinosa@tecnalia.com}
\affiliation{Department of Physics, University of the Basque Country UPV/EHU, 48080 Bilbao, Spain}
\affiliation{TECNALIA, Digital Unit Quantum Technologies Group. The Basque Country Technology Park. Astondo Bidea, Edificio 700, 48160 Derio, Spain}
\author{Lian-Ao Wu}
\email{lianao.wu@ehu.eus}
\affiliation{Department of Physics, University of the Basque Country UPV/EHU, 48080 Bilbao, Spain}
\affiliation{EHU Quantum Center, University of the Basque Country UPV/EHU, 48940 Leioa, Spain}
\affiliation{Ikerbasque, Basque Foundation for Science, 48013 Bilbao, Spain}

\date{\today}

\begin{abstract}
Quantum thermalization in contemporary quantum devices, in particular quantum computers, has recently attracted significant theoretical interest. Unusual thermalization processes, such as the Quantum Mpemba Effect (QME), have been explored theoretically. However, there is a shortage of experimental results due to the difficulty in preparing thermal states. In this paper, we propose a protocol to indirectly address this challenge. Moreover, we experimentally validate our protocol using IBM quantum devices, providing results for unusual relaxation in equidistant quenches as predicted for the IBM qubit. We also assess the formalism introduced for the QME, obtaining results consistent with the theoretical predictions. This demonstration underscores that our protocol can provide an alternative way of studying thermal states physics when their direct preparation may be too difficult.
\end{abstract}

\maketitle

\section{Introduction}
The current Noisy Intermediate-Scale Quantum (NISQ) era quantum computers are highly susceptible to decoherence \cite{nisq}. This poses a current limitation on the potential for promised exponential speedup of quantum algorithms over their classical counterparts. However, as will be demonstrated in this work, NISQ era devices serve as a controllable platform for studying and testing Open Quantum Systems (OQS) physics \cite{openquantumsystems}. In particular, we will illustrate that these devices can be employed for the study of quantum thermalization. Thermal states hold significant theoretical interest; for instance, the QME \cite{pnas}, an analog to the classical phenomenon where hot water freezes faster than cold water \cite{mpemba2021}, has been predicted for certain quantum systems \cite{Mpembareservoir, mpembaspin}. 

However, since thermal states are mixed states, they cannot be obtained from a pure initial state via unitary evolution. Preparing initial thermal states directly from experimental thermalization is highly impractical and would require precise control of the environment temperature at all times. Another possibile approach is to utilize quantum algorithms that prepare thermal states \cite{Wang_2021,sagastizabal2020variational,Guo_2023,verdon2019quantum, Ezzell_2023}. This algorithms typically rely on variational quantum algorithms (VQA's) and have been tested on NISQ devices \cite{sagastizabal2020variational,PRXQuantum.2.010317}.
This approach is of interest and utility for many quantum computation applications that require thermal state inializations, such as quantum simulation \cite{Childs_2018} or quantum machine learning \cite{Kieferov__2017}. Nonetheless, the study of quantum thermodynamics necessitates examining a large quantity of thermal states and the current algorithms have applicability and scalability issues \cite{Guo_2023}. Thus, the experimental cost of preparing many individual thermal states and manipulating them for the intended application is enormous. In this work, we address this issue by introducing a protocol for indirectly studying the afterward dynamics of initialized thermal states in any quantum system, provided that we can prepare pure eigenstates of a Hamiltonian. While our proposal cannot resolve the challenge of thermal state preparation, we will demonstrate that it significantly simplifies the task of studying the dynamics of initial thermal states. Therefore, it holds great practiality for basic research applications. 

We apply this protocol using two IBM quantum computers, experimentally testing the formalism for the QME in this system. Furthermore, we examine another unusal thermalization process known as Equidistant Quenches \cite{eqquenches, eqquenches2}, which contrasts the relaxation into equilibrium of a pure state with that of an initially equidistant thermal state.

The IBM Quantum Experience \cite{ibm} provides public access to superconducting qubits \cite{supereview}. These devices undergo a relaxation process into a thermal equilibrium state due to interaction with their environment, leading to a loss of information. To solve this problem, quantum error correction \cite{Roffe_2019} can be applied. However, such techniques require several additional qubits for correcting errors on a given qubit, making them impractical given that the current size of the largest quantum computers is on the order of $\sim$ 100 qubits. Alternatively, pulse sequences can be applied to protect the state from decoherence; for example, Dynamical Decoupling has been tested on IBM devices \cite{sym15010062}. Another approach involves designing the quantum algorithm to be inherently immune to external noise \cite{PhysRevA.106.012607,quiroz2024dynamically}.

The paper is structured as follows: In Section 2, we introduce the protocol and present the theoretical results essential for this work. Section 3 outlines the methodology employed in the demonstrations. The results obtained from the IBM devices are presented in Section 4, followed by the conclusion in Section 5.

Thermal states are defined by the Boltzmann distribution:

\begin{equation}\label{eq: thst1}
\rho_{\mathrm{th}}(T)=\frac{e^{-H / k_{\mathrm{B}} T}}{\operatorname{Tr}(e^{-H / k_{\mathrm{B}} T})}.
\end{equation}
Here, $H$ is the Hamiltonian, $k_{\mathrm{B}}$ is the Boltzmann constant, and $T$ is the temperature. In the eigenstates basis of the Hamiltonian, we can express the thermal state as:

\begin{equation}
\rho_{\mathrm{th}}(T)=\sum_{n} p_{n}(T) \ket{n}\bra{n}.
\end{equation}

The probability $p_{n}(T)$ for each eigenstate is given by:

\begin{equation}\label{probs}
p_{n}=\frac{e^{-E_{n} / k_{\mathrm{B}} T}}{\operatorname{Tr}(e^{-E_{n} / k_{\mathrm{B}} T})}.
\end{equation}
The dynamics of an OQS can be expressed using the Kraus Operator Sum Representation (KOSR) \cite{openquantumsystems} \cite{krausbook} as:

\begin{equation}
\begin{aligned}
\rho_{th}(T,t)&=\sum_{n,\alpha}p_{n}(T)K_{\alpha}(t,t_{0})\ket{n(t_{0})}\bra{n(t_{0})}K^{\dagger}_{\alpha}(t,t_{0})\\& =\sum_{n}p_{n}(T)\rho_{n}(t).
\end{aligned}
\end{equation}
Here, $K_{\alpha}$ represents the Kraus operators, satisfying the normalization condition $K_{\alpha}K^{\dagger}_{\alpha}=\mathbb{I}$. Since $\rho_{n}(t)$ are the time-dependent density matrices for each eigenstate, we can individually measure the evolution for each $\ket{n}$ and then add them with probability $p_{n}(T)$, obtaining the equivalent dynamics for any initial thermal state.

To model the IBM qubit, we will employ the same approximate Markovian master equation as in \cite{lidaribm}:

\begin{equation}\label{eq:qubitME}
\begin{aligned}
\dot{\rho}(t)=& -\frac{i}{\hbar}[H, \rho(t)]+\gamma\left(\sigma_{-} \rho(t) \sigma_{+}-\frac{1}{2}\left\{\sigma_{+} \sigma_{-}, \rho(t)\right\}\right)\\ &+e^{-\beta\hbar\omega}\gamma\left(\sigma_{+} \rho(t) \sigma_{-}-\frac{1}{2}\left\{\sigma_{-} \sigma_{+}, \rho(t)\right\}\right)\\ & +\gamma_{z}\left(\sigma_{z} \rho(t)\sigma_{z}-\rho(t)\right),
\end{aligned}
\end{equation}
where $\sigma_{\pm}$ and $\sigma_{z}$ denote the Pauli operators, $\beta$ is the inverse temperature, $\hbar$ is the Planck constant, $\gamma$ and $\gamma_{z}$ are the relaxation and dephasing rates, respectively, and $\omega$ is the qubit frequency. We will use the two-level Hamiltonian:

\begin{equation}\label{eq:qubitham}
H=-\frac{1}{2}\hbar\omega\sigma_{z}.
\end{equation}
Introducing this Hamiltonian in equation 5, the time evolution has an analytical solution that can be writen as:
\begin{equation}\label{eq:solution}
\begin{aligned}
\rho_{00}=&(\rho_{00}(0)-\rho_{00}^{eq})e^{-t\gamma(e^{\beta\hbar\omega}+1)}+\rho_{00}^{eq},\\
\rho_{01}=&\rho_{01}(0)e^{-\frac{1}{2}t(\gamma+e^{-\beta\hbar\omega}\gamma)}e^{-4\gamma_{z}t}e^{-i\omega t},\\
\rho_{10}=&\rho_{10}(0)e^{-\frac{1}{2}t(\gamma+e^{-\beta\hbar\omega}\gamma)}e^{-4\gamma_{z}t}e^{i\omega t},\\
\rho_{11}=&(\rho_{11}(0)-\rho_{11}^{eq})e^{-t\gamma(e^{\beta\hbar\omega}+1)}+\rho_{11}^{eq}.
\end{aligned}
\end{equation}
Where $\rho_{ij}$ are the reduced density matrix components and $\rho_{ij}^{eq}$ are the equilibrium state components. The derivation can be found in \cite{eqquenches}.  In this same article, the concept of Equidistant Quenches was introduced, where, under certain conditions, a thermal state and a pure state initially at the same distance from equilibrium relax differently. For a system following Equation \ref{eq:qubitME}, there is a critical value for the dephasing rate, $\gamma_{z,c}$:

\begin{equation}
\gamma_{z,c}\equiv\frac{\gamma(1+e^{-\beta\hbar\omega})}{8}.
\end{equation}

In the case of the studied IBM qubits, $\gamma_{z}>\gamma_{z,c}$. Consequently, we expect to observe an initial pure state relax faster than the corresponding equidistant thermal state \cite{eqquenches}. This result should be independent of the distance function chosen. To demonstrate this, we will compare the results using the trace distance and the KL divergence as well \cite{openquantumsystems}. Their definitions are:

\begin{equation} \label{eq:trdist}
D_{Tr}(\rho||\rho_{eq})=\frac{1}{2}\operatorname{Tr}(|\rho - \rho_{eq}|) ,
\end{equation}

\begin{equation} \label{eq:kldiv}
S(\rho||\sigma)\equiv \operatorname{Tr}(\rho \ln{\rho})-\operatorname{Tr}(\rho \ln{\sigma}),
\end{equation}
where $|A|=\sqrt{AA^{\dagger}}$.

For exploring the QME, we will use the formalism developed in \cite{pnas}. There, the following entropy distance is introduced:

\begin{equation}\label{eq:entrodist}
D_e\left[\vec{p} ; T_{eq}\right]=\sum_n\left(\frac{E_n \Delta p_n}{k_{B}T_{eq}}+p_n \ln p_n-p_n^{eq }\ln p_n^{eq}\right).
\end{equation}

It is important to note that this distance function is defined only for comparing thermal states. For two thermal states initially at $T_{eq}<T_{c}<T_{h}$, we always have $D_{e}[\vec{p}(T_{c},0),T_{eq}]<D_{e}[\vec{p}(T_{h},0),T_{eq}]$. The QME effect occurs if at any time $t$ there is a crossing and $D_{e}[\vec{p}(T_{c},t),T_{eq}]>D_{e}[\vec{p}(T_{h},t),T_{eq}]$. This would mean that the initial state at a higher temperature relaxes faster into equilibrium, an analog to the classical effect.

The ocurrence or absence of the effect in a Markovian system can be predicted from the form of the probability vector in terms of the eigenvectors of the master equation, that is,

\begin{equation}
\vec{p}(t)= \vec{p}_{eq}(T_{eq})+a_{2}e^{\lambda_{2}t}\vec{v}_{2}(t)+...+a_{n}e^{\lambda_{n}t}\vec{v}_{n}(t).
\end{equation}
Then, a system will present the QME \cite{pnas} if

\begin{equation}
    \lambda_{2}>\lambda_{3}\quad,\quad |a_{2}^{c}|>|a_{2}^{h}|.
\end{equation}
In the case of a system described by Equation (\ref{eq:qubitME}) the effect should not appear since we always have that $\lambda_{3}=0$ and $\lambda_{2}<0$ \cite{eqquenches}. Thus, for the IBM superconducting qubit, normal thermalization as defined in \cite{pnas} is expected for the initial thermal states. That is, the distance to equilibrium should be a monotonically increasing function of temperature at all times for $T>T_{eq}$. 
\section{Methods}
\begin{figure}
\centering
\includegraphics[width=\linewidth]{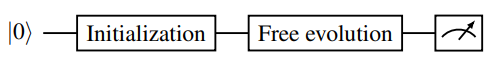}    
\caption{Scheme for the demonstrations, the particular gates corresponding to the initialization block varies depending on the state and the free evolution time is changed in each snapshot in order to observe the entire thermalization process.}
    \label{fig:1}
\end{figure}

In this work, we present two demonstrations, and a schematic representation of all executed circuits can be seen in Figure \ref{fig:1}. The first demonstration was conducted on qubit 0 of the 5-qubit \textit{ibmq\_belem} device. We prepared $\ket{0}$ and $\ket{1}$ along with two different superposition states, $\ket{\Psi_{1}}= \frac{2}{\sqrt{5}}\ket{0}+\frac{1}{\sqrt{5}}\ket{1}$ and $\ket{\Psi_{2}}= \frac{3}{\sqrt{10}}\ket{0}+\frac{1}{\sqrt{10}}\ket{1}$. We took 33 snapshots, each separated by 8 $\mu s$. The choice of this number is due to the limitation imposed by the IBM service, allowing only 100 different circuits in each usage of the quantum computer. We needed 3 circuits for each snapshot to perform Quantum State Tomography (QST) \cite{Cramer_2010} for reconstructing the entire density matrix for both pure states, considering their off-diagonal contributions. QST was implemented using the Qiskit Softwate Development Kit (SDK) \cite{Qiskit}.

For the second demonstration, on qubit 0 of the 5-qubit \textit{ibmq\_manila} computer, we prepared $\ket{0}$ and $\ket{1}$. We took 100 snapshots with a separation of 32 $\mu s$. As there were no off-diagonal terms contributing to any of the states in this case, QST was not needed. This allowed us to capture a higher number of snapshots. Diagonal density matrix elements were calculated by dividing the number of counts for each state by the total number of shots, $n_{\text{shots}}=8192$, in all circuits for both demonstrations.

In both demonstrations, thermal states were constructed as explained before, by adding the results for the pure initial eigenstates $\ket{0}$ and $\ket{1}$ at each time step with the corresponding ponderation given by Equation \ref{probs}. In Table \ref{tab:1}, we present the hardware calibration data provided by IBM at the time of the demonstration for \textit{ibmq\_belem} and \textit{ibmq\_manila}.

\begin{table}
\centering
\begin{tabular}{|c|c|c|c|}
\hline
\textrm{Device}&
\textrm{$\omega(GHz)$}&
\textrm{T1$(\mu s)$}&
\textrm{T2$(\mu s)$}\\
\hline
\textit{ibmq\_belem}	& 5.09			& 113.1 & 60.3 \\
\textit{ibmq\_manila}	& 4.9			& 178.9 & 90.6 \\
\hline
\end{tabular}
\caption{Calibration values at the time of the demonstration.\label{tab:1}}
\end{table}

\begin{figure}
    \centering
    \includegraphics[width=\linewidth]{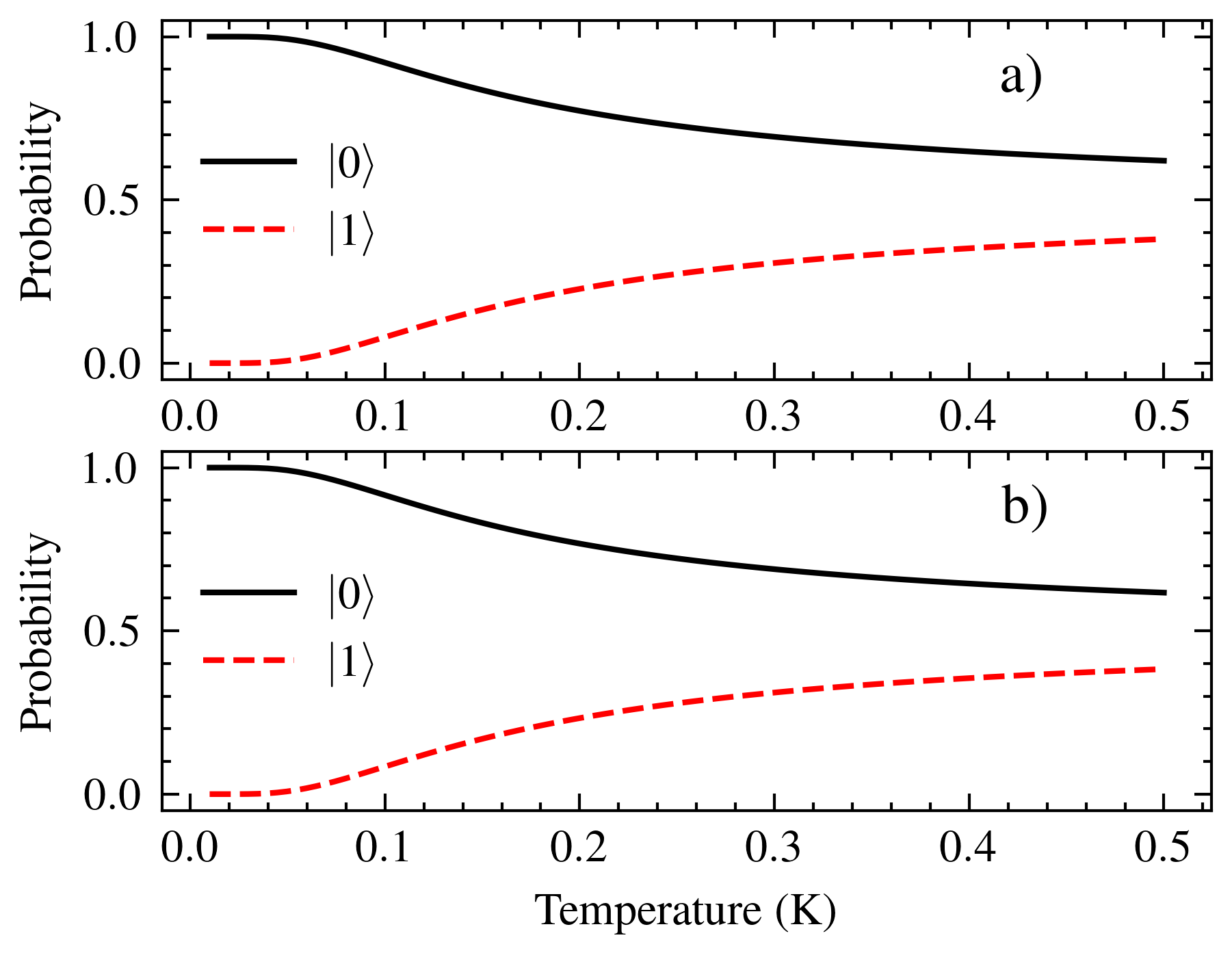}
    \caption{Thermal state eigenstates probabilities used for the thermal state construction as a function of temperature for (a) $\textit{ibmq\_belem}$  and (b) $\textit{ibmq\_manila}$.}
    \label{fig:2}
\end{figure}

The mixture probabilities used for constructing the thermal states are calculated using Equations (\ref{probs}) and (\ref{eq:qubitham}) along with the values from Table \ref{tab:1}. We present them in Figure \ref{fig:2}.

Communication with IBM services was performed using Qiskit. Data treatment was done in Python, utilizing the Matplotlib library for plotting and the Scipy package for numerical fitting of Equation \ref{eq:solution}. We left the equilibrium component as a free parameter and used the calibration values given in Table I for the calculation of the dephasing and relaxation rates.
\section{Results}
\begin{figure}
\centering
    \includegraphics[width=0.8\linewidth]{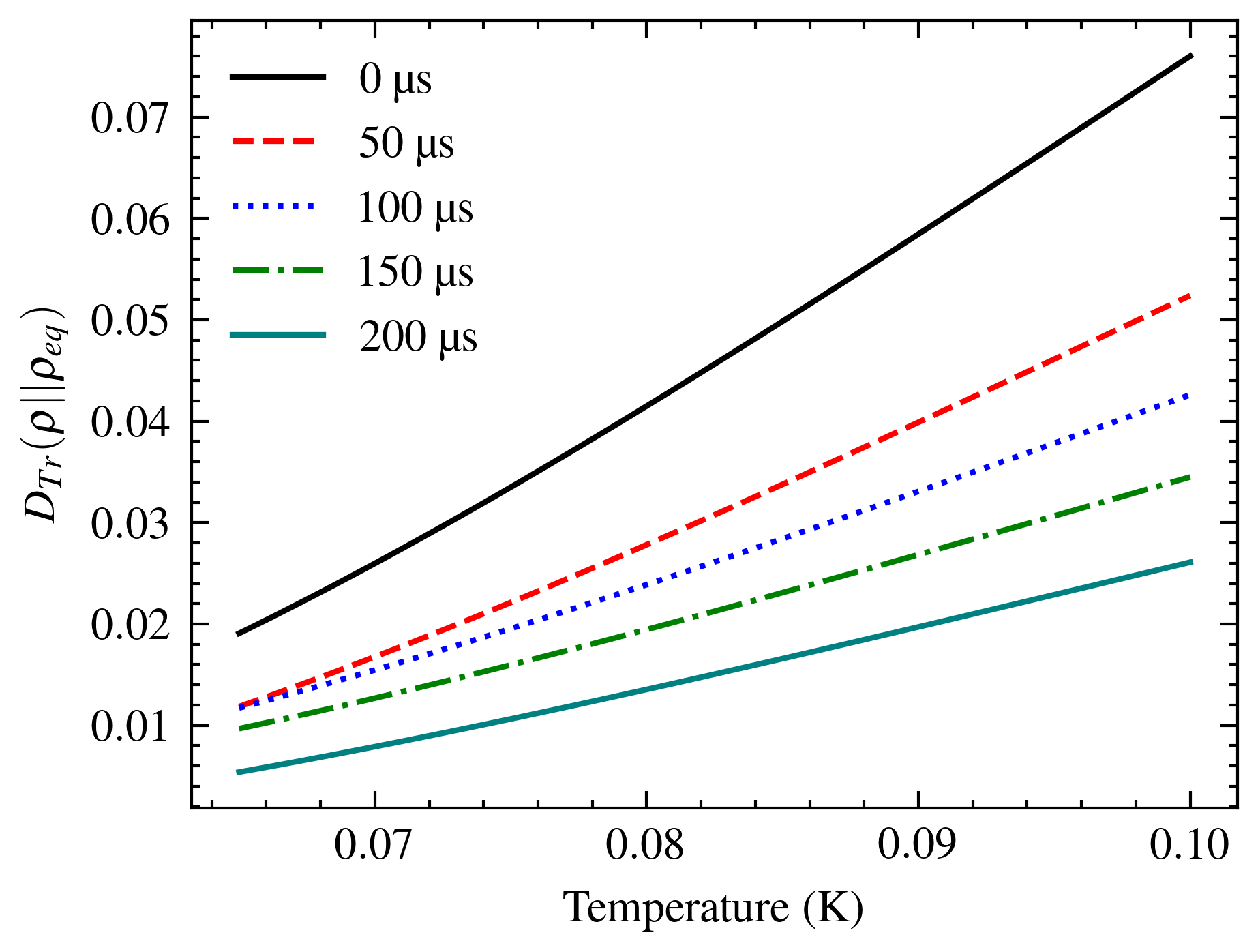}
    \caption{Trace distance as a function of temperature for thermal states on \textit{ibmq\_manila} at different times.}
    \label{fig:3}
\end{figure}

\begin{figure}
\centering
    \includegraphics[width=0.8\linewidth]{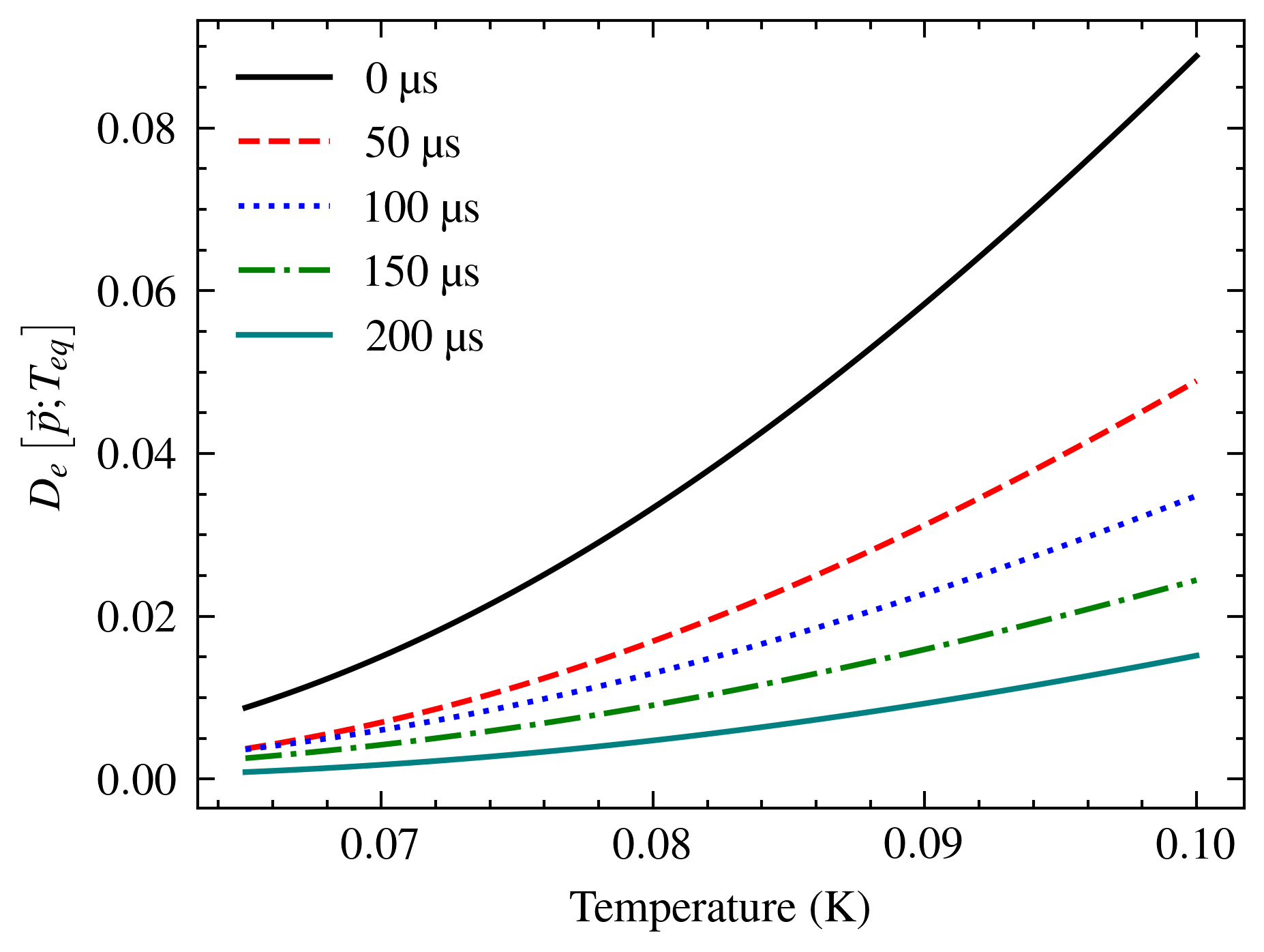}
    \caption{Entropic distance as a function of temperature for thermal states on \textit{ibmq\_manila} at different times.}
    \label{fig:4}
\end{figure}

For thermal states on \textit{ibmq\_manila}, in Figure \ref{fig:3} and Figure \ref{fig:4} it is shown that, for both distance functions, we observe a monotonic increase with temperature at all times. This indicates that we do not observe the QME, which is consistent with the theory that does not predict QME to occur in a system described by Equation (\ref{eq:qubitME}).

\begin{figure}
\centering
    \includegraphics[width=0.8\linewidth]{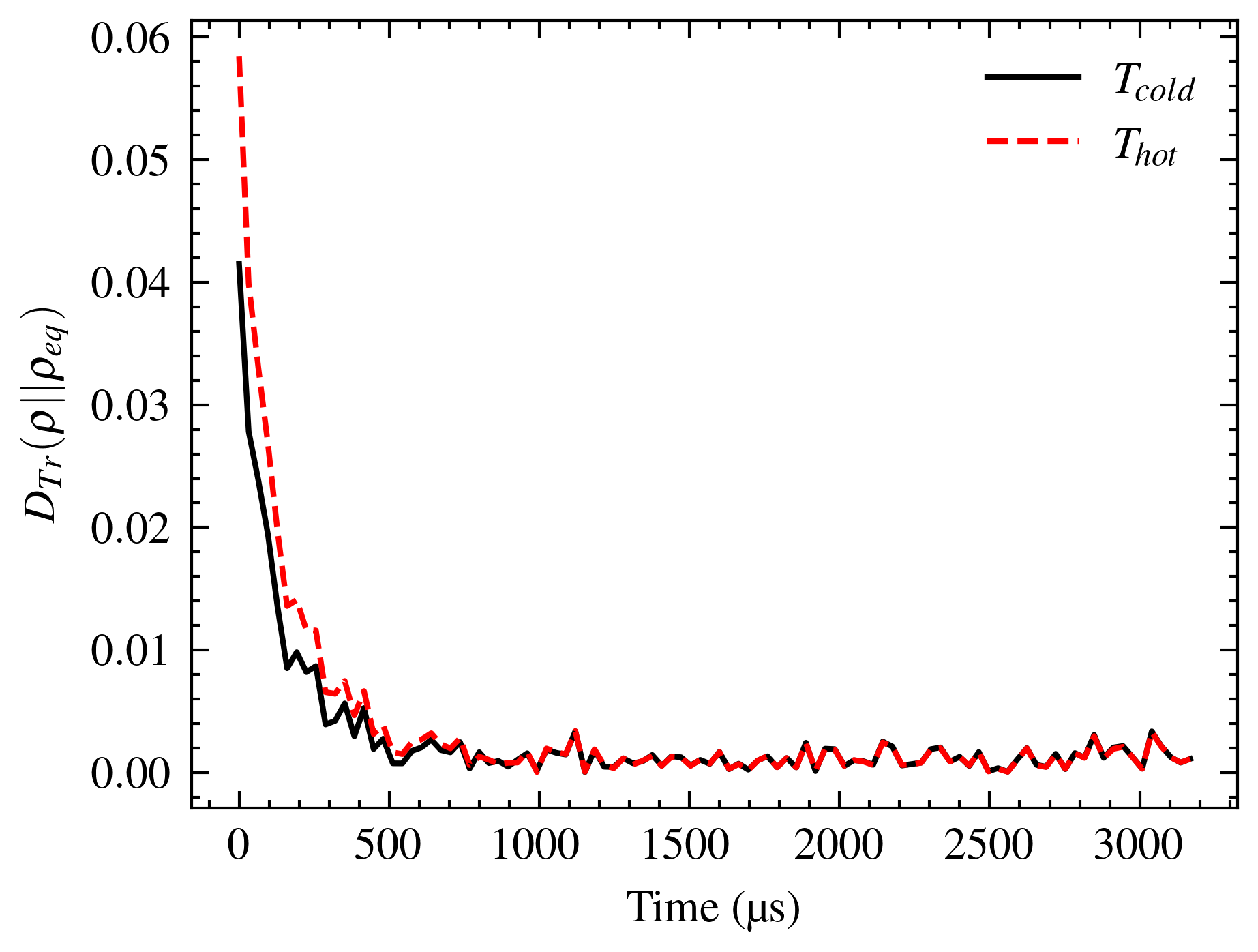}
    \caption{Trace distance as a function of time for two different initial thermal states on \textit{ibmq\_manila}, $T_{cold}= 80$ mK and $T_{hot}= 90$ mK.}
    \label{fig:5}
\end{figure}

\begin{figure}
\centering
    \includegraphics[width=0.8\linewidth]{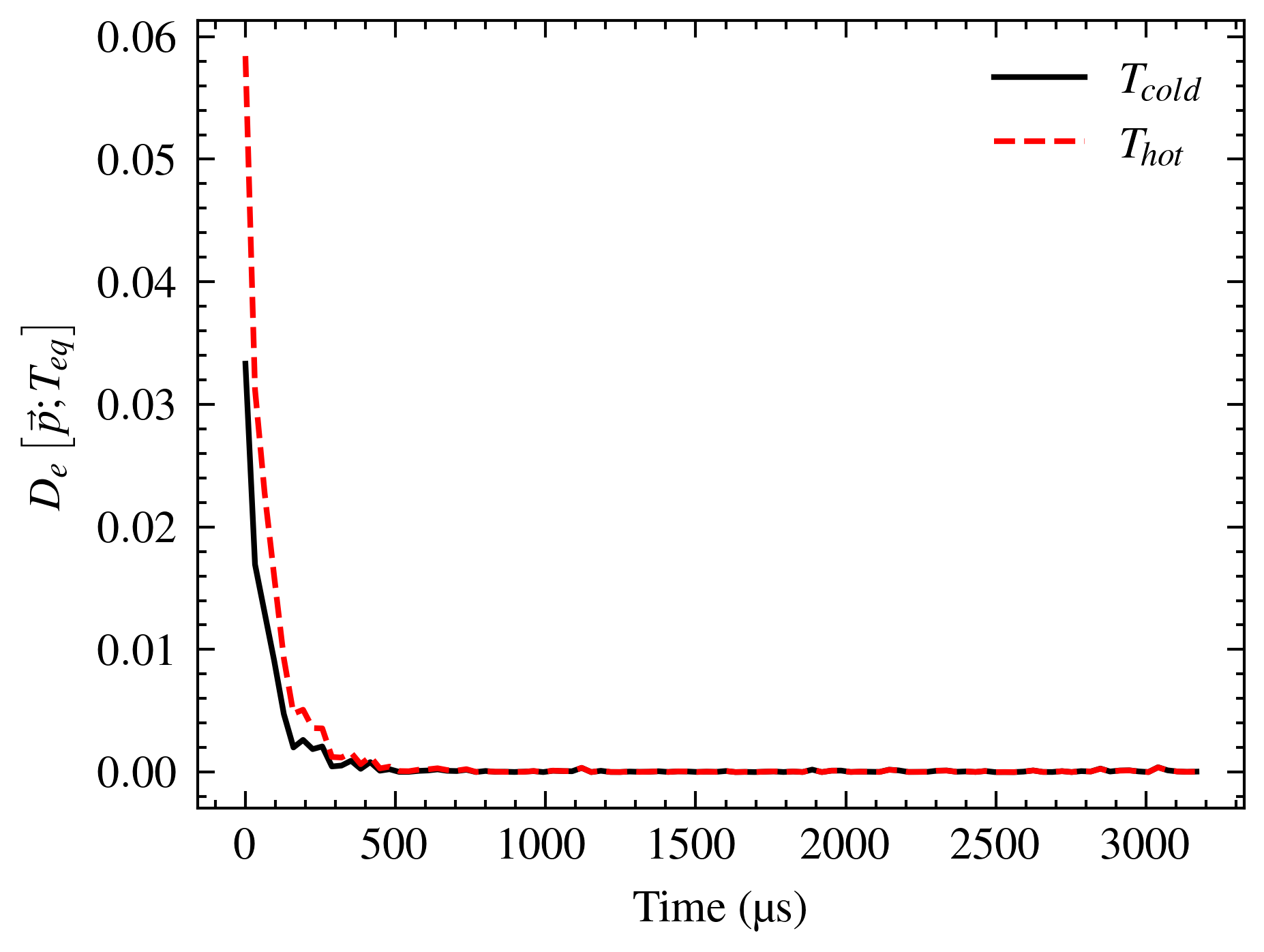}
    \caption{(\textbf{a}) Entropic distance as a function of time for two different initial thermal states on \textit{ibmq\_manila}, $T_{cold}= 80$ mK and $T_{hot}= 90$ mK.}
    \label{fig:6}
\end{figure}

In Figures \ref{fig:5} and \ref{fig:6}, we compare the relaxation of two different initial thermal states with both distance functions. Once again, there is no crossing between the curves. The equilibrium temperature found was $T_{eq}=58$ mK.

\begin{figure*}
\centering
\includegraphics[width=0.9\linewidth]{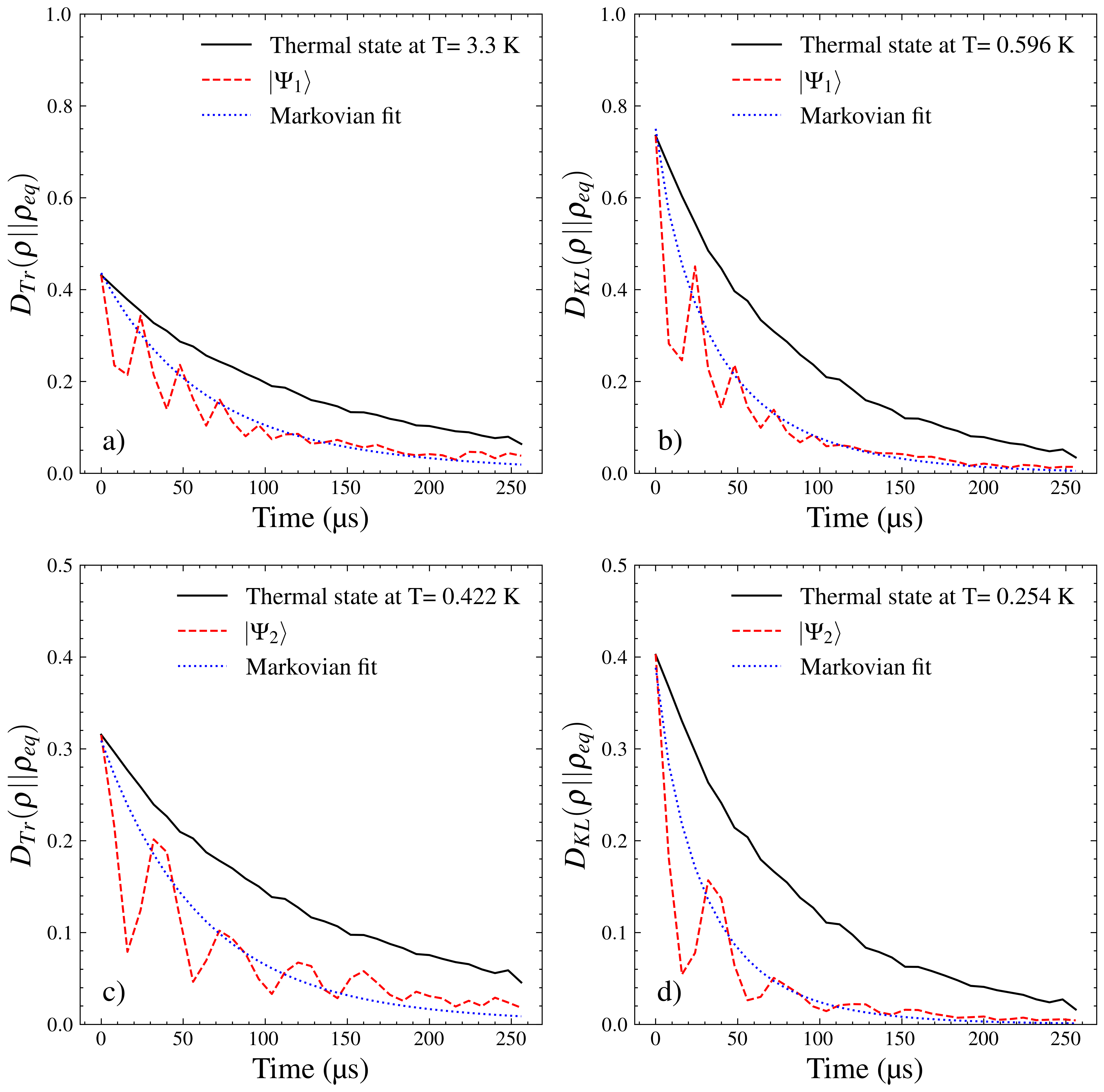}
\caption{(\textbf{a}) Time evolution of the trace distance for $\ket{\Psi_{1}}$, compared to an initially equidistant thermal state and the Markovian equation prediction. (\textbf{b}) Same as (\textbf{a}) but using the KL divergence. (\textbf{c}-\textbf{d}) Same as (\textbf{a}-\textbf{b}) respectively for the state $\ket{\Psi_{2}}$ .\label{fig:7}}
\end{figure*}

Finally, in Figure \ref{fig:7}, we present the results for the \textit{ibmq\_belem} device. We observe, for both states using both distance functions, a faster relaxation into equilibrium than the corresponding thermal state, as predicted in \cite{eqquenches}. The oscillations in the distance for the pure states are a clear sign of Non-Markovianity, as previously reported on IBM devices \cite{lidaribm}. Conversely, the thermal states exhibit Markovian behavior. A possible reason for this difference is that the dynamics of pure states are affected by T2, while for the thermal state, only T1 dictates the timescale of the evolution. However, we note that the expected effect is observed in all cases, even considering the Non-Markovianity of the system.

As observed, the equidistant thermal state does depend on the distance function, illustrating the fact that temperature in quantum systems is ill-defined. The equilibrium state was determined by fitting the time evolution of $\ket{0}$, providing an estimate for the computer temperature of $T_{eq}=67$ mK.
\section{Discussion}
In this work, we have introduced and successfully tested a powerful protocol for exploring the physics of thermal states in quantum systems, which, in principle, could be extended to other quantum systems. This significantly simplifies the challenge of obtaining experimental results in the field of quantum thermodynamics. More significantly, we can  demonstrate quantum thermalization from thermal initial states directly. We have also provided experimental evidence of coherent states relaxing faster into equilibrium than equidistant thermal states, as predicted in \cite{eqquenches} and \cite{eqquenches2} theoretically. Furthermore, we have identified strong non-Markovian effects on the IBMQ devices, in line with \cite{lidaribm}, offering additional experimental insights into this matter. These findings are crucial to consider when implementing decoherence control techniques \cite{sym15010062}. We observed that thermal states on the quantum computer follow a normal, Markovian relaxation into equilibrium, suggesting that Markovian equations may be sufficient for describing this type of states in this particular system under longer time limit.

However, it is important to highlight the limitations and weaknesses of the proposed protocol. Firstly, despite the formal equivalence between eigenstates and thermal state evolution, thermal states are not actually prepared on the quantum system. Therefore, our protocol may not be applicable for quantum algorithms that utilize thermal states as a computational tool. Additionally, ensuring identical interactions with the environment for each eigenstate evolution may be necessary. We have addressed this by running the different circuits as closely in time as possible. Moreover, this protocol needs to be further studied on different quantum systems with higher complexity. In particular, it would be desirable to apply our protocol on a highly controlled environment in order to further validate the protocol.

Through simple demonstrations, we have indirectly tested numerous results of quantum thermodynamics that would otherwise require a challenging experimental setup. This underscores the potential utility of NISQ quantum computers not only as calculating devices but also as systems for studying and testing Open Quantum Systems physics.
\acknowledgements

We would like to thank Dr Dawei Luo for discussions. We gratefully acknowledge supports from the Basque Country Government (Grant No. IT1470- 22) and Grant No. PGC2018-101355-B-I00 funded by MCIN/AEI/10.13039/501100011033. This project has also received support from the Spanish Ministry for Digital Transformation and of Civil Service of the Spanish Government through the QUANTUM ENIA project call - Quantum Spain, EU through the Recovery, Transformation and Resilience Plan – NextGenerationEU within the framework of the Digital Spain 2026.

\end{document}